\documentclass[10pt,conference]{IEEEtran}

\usepackage[table, dvipsnames]{xcolor}
\usepackage{cite}
\usepackage{amsmath,amssymb,amsfonts}
\usepackage{algorithmic}
\usepackage{graphicx}
\usepackage{textcomp}
\usepackage{hyperref}
\usepackage{booktabs}
\usepackage{nicematrix}
\usepackage{multirow}
\usepackage{tcolorbox}
\usepackage{subcaption}
 
\newcommand{\tool}{\textsc{Code-DiTing}}

\IEEEoverridecommandlockouts
\def\BibTeX{{\rm B\kern-.05em{\sc i\kern-.025em b}\kern-.08em
    T\kern-.1667em\lower.7ex\hbox{E}\kern-.125emX}}
\begin{document}


\title{{\tool}: A Reasoning-Based Metric for Functional Alignment in Code Evaluation}


\author{\IEEEauthorblockN{Guang Yang\IEEEauthorrefmark{2},  
Yu Zhou\IEEEauthorrefmark{2}\IEEEauthorrefmark{1}, 
Xiang Chen\IEEEauthorrefmark{3},  
Wei Zheng\IEEEauthorrefmark{4},  
Xing Hu\IEEEauthorrefmark{5},
Xin Zhou\IEEEauthorrefmark{6},
David Lo\IEEEauthorrefmark{6},
Taolue Chen\IEEEauthorrefmark{8}\IEEEauthorrefmark{1}}
\IEEEauthorblockA{
\IEEEauthorrefmark{2}\textit{College of Computer Science and Technology}, \textit{Nanjing University of Aeronautics and Astronautics}, China\\
\IEEEauthorrefmark{3}\textit{School of Artificial Intelligence and Computer Science}, \textit{Nantong University}, China\\
\IEEEauthorrefmark{4}\textit{School of Software}, \textit{Northwestern Polytechnical University}, China\\
\IEEEauthorrefmark{5}\textit{School of Software Technology},\\
\textit{Zhejiang University}, China\\
\IEEEauthorrefmark{6}\textit{School of Computing and Information Systems},\\
\textit{Singapore Management University}, Singapore\\
\IEEEauthorrefmark{8}\textit{School of Computing and Mathematical Sciences},\\
\textit{Birkbeck, University of London}, UK\\
Email: novelyg@outlook.com, zhouyu@nuaa.edu.cn, xchencs@ntu.edu.cn, \\
wzheng@nwpu.edu.cn, xinghu@zju.edu.cn, xinzhou.2020@phdcs.smu.edu.sg, \\
davidlo@smu.edu.sg, t.chen@bbk.ac.uk}
}

\maketitle

\begingroup
\renewcommand{\thefootnote}{}
\footnotetext[1]{\IEEEauthorrefmark{1} Yu Zhou and Taolue Chen are the corresponding authors.}
\endgroup

\begin{abstract}
Trustworthy evaluation methods for code snippets play a crucial role in neural code generation. 
Traditional 
methods, which either rely on reference solutions or require executable test cases, have inherent limitation in 
flexibility and scalability.
The recent LLM-as-Judge methodology offers a promising alternative by directly evaluating functional consistency between the problem description and the generated code.
To systematically understand the landscape of these LLM-as-Judge methods, we conduct a comprehensive empirical study across three diverse datasets. Our investigation reveals the pros and cons of two categories of LLM-as-Judge methods: 
the methods based on general foundation models can achieve good performance but require complex prompts and lack explainability, while the methods based on reasoning foundation models provide better explainability with simpler prompts but demand substantial computational resources due to their large parameter sizes.

To address these limitations, we propose {\tool}, a novel code evaluation method that balances accuracy, efficiency and explainability. 
We develop a data distillation framework that effectively transfers reasoning capabilities from DeepSeek-R1-671B to our 
{\tool} 1.5B and 7B models, significantly enhancing evaluation explainability and reducing the computational cost.
With the majority vote strategy in the inference process, {\tool} 1.5B outperforms all models with the same magnitude of parameters and achieves performance which would normally exhibit 
in a model with 5 times of parameter scale. 
{\tool} 7B surpasses GPT-4o and DeepSeek-V3 671B, even though it only uses 1\% of the parameter volume of these large models.
Further experiments show that {\tool} is robust to preference leakage and can serve as a promising alternative for code  evaluation.
\end{abstract}


\section{Introduction}

Large Language Models (LLMs) have emerged as a fundamental tool in modern software development~\cite{he2024llm, shi2024efficient, hou2024large}, demonstrating exceptional language understanding and generation capabilities.
Their application has shown remarkable potential across various software engineering tasks \cite{fan2024exploring, yang2024chain}, particularly in code generation\cite{yang2025empirical}.
However, as LLMs are increasingly deployed, evaluating the correctness of generated code remains a significant challenge \cite{wang2023review, zheng2024towards}, primarily because multiple correct or semantically equivalent solutions \cite{yang2025assessing} may exist for a  given programming problem.

Traditional evaluation metrics, which are either reference-based or test-based, have been widely adopted. 
However, these metrics suffer from inherent limitations. 
Reference-based metrics (e.g., BLEU~\cite{papineni2002bleu}, ROUGE~\cite{lin2004rouge} and ChrF~\cite{popovic2015chrf}) depend on high-quality reference code and frequently penalize implementations that are correct but diverge from them. 
Test-based metrics (e.g., Pass@k~\cite{chen2021evaluating}) require careful manual design of comprehensive test cases that cover edge cases, along with secure environments for code execution.
Another evaluation method is human evaluation~\cite{yang2023exploitgen}, which is accurate yet expensive, as it involves multiple domain experts who directly assess the correctness of generated artifacts. 
Importantly, this method is prohibitively labor-intensive and time-consuming, rendering it impractical for large-scale assessments.
These constraints significantly limit the flexibility and scalability of human evaluation for code generation evaluation~\cite{takaichi2022nlp, naik2024limitations}.

Recent advancements in LLMs have catalyzed the development of LLM-as-Judge methods~\cite{li2024llms, gu2024survey, li2024generation}, which directly evaluate the functional consistency between problem descriptions and generated code. 
These methods offer a promising alternative to traditional evaluation method~\cite{zheng2023judging}. 
However, with the rapid proliferation of LLM-as-Judge methods, there remains considerable uncertainty regarding their performance in code generation evaluation and it is far from clear which method delivers optimal results~\cite{yu2024fight}.

\smallskip
\noindent \textbf{Empirical study.} 
We first conduct a 
large-scale empirical study to systematically compare different LLM-as-Judge methods in code generation evaluation. 
Specifically, we classify existing LLM-as-Judge methods into two categories, i.e., methods based on general models (e.g., GPT-3.5-turbo and GPT-4o) and  methods based on reasoning-focused models (e.g., DeepSeek-R1\cite{guo2025deepseek}). 
To ensure the comprehensive evaluation, we curate three datasets (i.e., HumanEval-Judge, MBPP-Judge and BigCodeBench-Judge) as new benchmarks for evaluating the effectiveness of LLM-as-Judge methods in code generation evaluation.
Our findings indicate that, while these methods generally perform well, they exhibit significant discrepancy across various dimensions. 
In particular, the former 
requires elaborate prompts and lacks explainability, whereas the latter provides enhanced explainability with simpler prompts but demands substantial computational resources due to their parameter sizes.

\smallskip
\noindent \textbf{Our methods.} To address these limitations and advance the state of code generation evaluation, we propose a novel code evaluation method that effectively balances accuracy, efficiency, and explainability. 
We name it {\tool}\footnote{The name is from Chinese classic \textit{Journey to the West}, reflecting the model's goal to accurately discern the correctness of code implementations, just as the mythical creature distinguishes truth from falsehood.}.

To reduce the computational cost, we develop a data distillation framework that transfers reasoning capabilities from the powerful DeepSeek-R1-671B model to our more compact {\tool} model, available in 1.5B and 7B parameter sizes. 
Through this process, we construct a high-quality dataset \textsc{CodeJudge-17K} consisting of 17,000 carefully curated samples with reasoning paths. 
This method not only enhances the explainability of the evaluation but also makes the reasoning process more accessible and comprehensible.
To further enhance performance, the {\tool} models employ PiSSA~\cite{meng2024pissa} technique for model training and the majority vote strategy during inference.

Experimental results demonstrate that {\tool} 1.5B outperforms all models of comparable parameter magnitude and achieves performance equivalent to models with five times the parameter count. 
Notably, {\tool} 7B surpasses even large-scale models such as GPT-4o and DeepSeek-V3 671B~\cite{liu2024deepseek}, despite utilizing only 1\% of their parameter volume. 
Our ablation studies reveal that all components of {\tool} are essential for its superior performance.
In addition, we demonstrate that {\tool} is robust to preference leakage~\cite{li2025preference}, where evaluation models show bias toward code produced by same series of architectures, a common issue in LLM-as-Judge methods.
These findings establish {\tool} as a promising alternative for code generation evaluation, representing a significant advancement in the field.

\smallskip
\noindent \textbf{Summary of contributions.} 
\begin{itemize}
    \item We curate three datasets (i.e., HumanEval-Judge, MBPP-Judge and BigCodeBench-Judge) as benchmark  for the empirical study. In addition, we introduce a new dataset \textsc{CodeJudge-17K} designed for training purposes.
    \item We design and carry out a large-scale empirical study to systematically compare different LLM-as-Judge methods in code generation evaluation.
    \item We propose {\tool}, a novel code evaluation method that effectively balances accuracy, efficiency and explainability.
    \item We conduct extensive experiments to evaluate the performance of {\tool} on different scenarios, including performance comparisons, ablation studies and analyses of preference leakage.
\end{itemize}


To facilitate reproducibility, experimental data and model weights are released at \url{https://github.com/Code-DiTing}.

\section{Background}
\label{sec:background}

\subsection{Problem Formulation}
We formally define the \emph{code generation evaluation problem} as follows. 
Let $\mathcal{X}$ be the space of problem descriptions, $\mathcal{Y}$ be the space of code implementations, $\mathcal{R}$ be the space of reference implementations and $\mathcal{T}$ be the space of test case sets. 

Given a problem description $x \in \mathcal{X}$, a code generation model $M: \mathcal{X} \rightarrow \mathcal{Y}$ produces code $y = M(x)$.
The evaluation function $\mathcal{F}: \mathcal{X} \times \mathcal{Y} \times \mathcal{R} \times \mathcal{T} \rightarrow \{0, 1\}$ determines the functional correctness of $y$ with respect to $x$. 
Formally, 
\begin{equation}
\mathcal{F}(x, y, r, T) = 
\begin{cases}
1, & \text{if } y \text{ is functionally correct}\\
0, & \text{otherwise}
\end{cases}
\end{equation}
where $r \in \mathcal{R} \cup \{\bot\}$ is an (optional) reference implementation ($r=\bot$ means that $r$ is not provided) and $T \in \mathcal{T} \cup \{\bot\}$ is an (optional) set of test cases ($T=\bot$ means that $T$ is not provided). 

Based on the availability of $r$ or $T$, the existing code generation evaluation methods can be categorized into: reference-based, test-based, and reference-and-Test-free evaluation.
Table~\ref{tab:metrics} summarizes a comparison of code generation evaluation metrics used in various methods.

\begin{table}[t]
\caption{Comparison of Code Generation Evaluation Metrics, where Func. means functional correctness, Auto. means automatic evaluation, Expl. means explainability and Open. means using open-source models. $\checkmark$ denotes applicable, $\times$ denotes not applicable  and $\circ$ denotes optional.}
\label{tab:metrics}
\rowcolors{4}{gray!15}{}
\centering
\resizebox{0.48\textwidth}{!}{
\begin{tabular}{lcccccc}
\toprule
\textbf{Metric} & \multicolumn{2}{c}{\textbf{Category}} & \multicolumn{4}{c}{\textbf{Characteristics}} \\
\cmidrule(lr){2-3} \cmidrule(lr){4-7}
    & \textbf{Ref} & \textbf{Test} & \textbf{Func.} & \textbf{Auto.} & \textbf{Expl.} & \textbf{Open.} \\
\midrule
BLEU~\cite{papineni2002bleu} & $\checkmark$ & $\times$ & $\times$ & $\checkmark$ & $\times$ & $\checkmark$\\
Rouge~\cite{lin2004rouge} & $\checkmark$ & $\times$ & $\times$ & $\checkmark$ & $\times$ & $\checkmark$\\
ChrF~\cite{popovic2015chrf} & $\checkmark$ & $\times$ & $\times$ & $\checkmark$ & $\times$& $\checkmark$\\
EM~\cite{liguori2023evaluates} & $\checkmark$ & $\times$ & $\times$ & $\checkmark$ & $\times$& $\checkmark$\\
ED~\cite{liguori2023evaluates} & $\checkmark$ & $\times$ & $\times$ & $\checkmark$ & $\times$& $\checkmark$\\
CrystalBLEU~\cite{eghbali2022crystalbleu} & $\checkmark$ & $\times$ & $\times$ & $\checkmark$ & $\times$& $\checkmark$\\
CodeBLEU~\cite{ren2020codebleu} & $\checkmark$ & $\times$ & $\times$ & $\checkmark$ & $\times$& $\checkmark$\\
CodeBERTScore~\cite{zhou2023codebertscore} & $\checkmark$ & $\times$ & $\times$ & $\checkmark$ & $\times$ & $\checkmark$\\
CodeScore~\cite{dong2025codescore} & $\checkmark$ & $\times$ & $\circ
$ & $\checkmark$ & $\times$ & $\checkmark$\\
CodeScore-R~\cite{yang2024codescore} & $\checkmark$ & $\times$ & $\checkmark$ & $\checkmark$ & $\times$ & $\checkmark$\\
\midrule
Human Study~\cite{yang2023exploitgen} & $\circ$ & $\times$ & $\checkmark$ & $\times$ & $\checkmark$ & $\checkmark$\\
Pass@k~\cite{chen2021evaluating} & $\times$ & $\checkmark$ & $\checkmark$ & $\checkmark$ & $\checkmark$ & $\checkmark$ \\
\midrule
ICE-Score~\cite{zhuo2024ice} & $\circ$ & $\times$ & $\checkmark$ & $\checkmark$ & $\times$ & $\times$\\
CodeJudge~\cite{tong2024codejudge} & $\circ$ & $\times$ & $\checkmark$ & $\checkmark$ & $\times$ & $\times$\\
\midrule
{\tool} & $\times$ & $\times$ & $\checkmark$ & $\checkmark$ & $\checkmark$ & $\checkmark$\\
\bottomrule
\end{tabular}
}
\end{table}

\subsection{Reference-Based Evaluation ($r \neq \bot$)}
Reference-based methods compute the similarity between $y$ and $r$, based on metrics ranging from token-based metrics (e.g., BLEU~\cite{papineni2002bleu}, ChrF~\cite{popovic2015chrf}) to semantics-aware ones (e.g., CodeBLEU~\cite{ren2020codebleu}, CodeBERTScore~\cite{zhou2023codebertscore}). 

Token-based metrics are limited to the n-gram lexical similarity computation and ignore potential semantic information in the code. These metrics originate from, e.g., machine translation and text summarization, including BLEU~\cite{papineni2002bleu}, ROUGE~\cite{lin2004rouge} and ChrF~\cite{popovic2015chrf}. 
Additionally, exact match (EM) metrics are widely used in code synthesis.
Eghbali et al.~\cite{eghbali2022crystalbleu} proposed the CrystalBLEU metric to enhance evaluation accuracy by excluding common n-grams that inflate BLEU scores due to verbose syntax and coding conventions.
Furthermore, Liguori et al.~\cite{liguori2023evaluates} argued that edit distance (ED) better measures code similarity compared to other token-based metrics.

Semantics-based metrics consider the syntactic structure, data flow information and potential semantic information of code. Ren et al.~\cite{ren2020codebleu} proposed CodeBLEU, which injects code syntax through AST and code semantics through data flow.
Dong et al.~\cite{dong2025codescore} proposed CodeScore, which conducts supervised learning on datasets with test cases to perform functional evaluation of code synthesis. 
Zhou et al.~\cite{zhou2023codebertscore} proposed CodeBERTScore, which uses CodeBERT to performs contextual encoding of reference and predicted code to calculate similarity scores between each token.
Yang et al.~\cite{yang2024codescore} proposed CodeScore-R based on UniXcoder and contrastive learning, which employs sketch processing, syntax transformation and mutation testing to improve the robustness of metric.

Nevertheless, these methods cannot directly assess functional correctness, require high-quality reference code collection, and penalize correct but divergent implementations.

\subsection{Test-Based Evaluation ($T \neq \bot$)}
Test-based methods 
~\cite{chen2021evaluating} execute code against test cases $T$ to assess functional correctness. 
The widely-adopted pass@k metric is defined as 
\begin{equation*}
\text{pass@k} = \mathbb{E}_x \left[ 1 - \frac{\binom{n-c}{k}}{\binom{n}{k}} \right]
\end{equation*}
where $n$ (resp.\ $c$) is the total (resp.\ correct) number of samples for the problem $x$.
This metric has become standard in evaluating code generation models.
    
Despite its popularity, pass@k requires human experts for designing high-quality test cases, and demands secure execution environments to prevent malicious code execution.

\subsection{Reference-and-Test-Free Evaluation ($r = \bot$ and $T = \bot$)}
When neither reference implementations nor test cases are available, evaluation typically relies on either human evaluation or LLM-as-judge methods~\cite{paul2024benchmarks, he2025code}. 
Human evaluation, while accurate, is prohibitively expensive and time-consuming for large-scale assessments. 
Yang et al.~\cite{yang2023exploitgen} proposed a sampling-based method applied in small-scale empirical studies, which can obtain high-quality assessments of predicted code within acceptable time and resource constraints.

Recent LLM-as-judge methods leverage large language models to directly evaluate the functional consistency between problem descriptions and generated code. 
Zhuo et al.~\cite{zhuo2024ice} proposed ICE-Score, which pioneered the use of GPT-3.5 as a judge to evaluate code generation model performance through carefully crafted prompt engineering.
Tong et al.~\cite{tong2024codejudge} introduced CodeJudge, which not only utilizes GPT-3.5 but also explores smaller open-source models as judges, employing a two-stage prompt engineering method for evaluation.

While promising, these methods generally require complex prompt engineering, rely on proprietary closed-source models, and lack 
explanations for their judgments.
In contrast, 
we aim to provide a simple and explainable evaluation method that requires neither reference implementations nor test cases, which can balance accuracy, efficiency, and explainability.

\section{Empirical Study}
\label{sec:empirical}
In this section, we conduct an empirical study to explore the 
existing LLM-as-judge methods to code generation evaluation  and analyze the various factors on their effectiveness.

\subsection{Experiment Setup}

\noindent\textbf{Code Generation Datasets.}
To comprehensively evaluate LLM-as-judge methods, establishing accurate and diverse benchmarks is a crucial first step. 
We select three diverse and widely adopted datasets that faithfully simulate real-world code generation scenarios.
Our dataset selection is guided by two principles:
(1) To ensure accurate assessment of semantic correctness, we prioritize datasets with exceptional test case quality and quantity, specifically targeting those with test coverage approaching 100\%;
(2) Beyond algorithm-centric problems, datasets need to encompass a wide range of libraries and function call patterns typical in professional software development, enabling thorough evaluation of LLM-as-judge methods across varied programming contexts.

As a result, we select the following datasets:
\begin{itemize}
    \item \textbf{HumanEval-plus}~\cite{evalplus} is an enhanced variant of the HumanEval benchmark that addresses fundamental ground-truth issues in the original dataset (including unhandled edge cases, logical errors and performance limitations). It expands the test coverage from an average of 9.6 to 764.1 test cases per problem, incorporating more challenging edge cases and complex functionalities to ensure rigorous and comprehensive evaluation.
    \item \textbf{MBPP-plus}~\cite{evalplus} applies similar enhancement techniques to the MBPP benchmark, resulting in a test suite 35 times larger than the original dataset. 
    \item \textbf{BigCodeBench}~\cite{zhuo2024bigcodebench} specifically targets real-world software development scenarios by incorporating diverse libraries and complex function call patterns. It comprises 1,140 function-level tasks that challenge LLMs to interpret instructions and orchestrate multiple function calls across 139 different libraries. Each programming task is validated through an average of 5.6 carefully designed test cases, achieving a mean branch coverage of 99\%.
\end{itemize}


\noindent\textbf{Data Sampling.}
With the chosen benchmark datasets, we proceed to sample code generated by various LLMs. 
We employ different models of varying sizes: Qwen2.5Coder (1.5B/7B)~\cite{hui2024qwen2} and DeepSeekCoder (1.3B/6.7B)~\cite{guo2024deepseek} to ensure diversity in the generated solutions.  
Using multiple models not only enhances the diversity of our dataset but also allows us to evaluate the robustness of LLM-as-Judge methods across different code generation patterns and qualities.

During the data processing phase, we extract natural language problem descriptions and corresponding code implementations from the generated samples through rigorous data cleaning and deduplication processes. 
Additionally, we remove code comments to enhance conciseness and focus the evaluation on functional implementation rather than documentation.

\noindent\textbf{Data Labeling.}
(1) Automatic Labeling. We utilize test cases from the existing datasets to automatically label code samples. 
Functional correctness is determined using the pass@1 metric, serving as the ground-truth for evaluation.
(2) Manual Verification. 
To address potential mislabeling from expanded test cases in HumanEval-plus/MBPP-plus, three authors independently review samples that passed original benchmarks but failed enhanced test suites. 
Labels are assigned directly when judgments align, or through discussion when opinions differ, ensuring high-quality ground-truth labels.

We hence curate three datasets: HumanEval-Judge (640 samples), MBPP-Judge (1,512 samples) and BigCodeBench-Judge (800 samples). Detailed statistics, including class distributions, are provided in Table \ref{tab:dataset_statistics}.

\begin{table}[t]
	\centering
	\caption{Sample Statistics for HumanEval-Judge, MBPP-Judge and BigCodeBench-Judge Datasets}
	\label{tab:dataset_statistics}
	\begin{tabular}{lccc}
		\toprule
		\textbf{Dataset} & \textbf{Samples} & \textbf{\#Positive} & \textbf{\#Negative} \\
		\midrule
		HumanEval-Judge & 640 & 480 (75.0\%) & 160 (25.0\%) \\
		MBPP-Judge & 1,512 & 997 (65.9\%) & 515 (34.1\%) \\
		BigCodeBench-Judge & 800 & 321 (40.1\%) & 479 (59.9\%) \\
		\bottomrule
	\end{tabular}
\end{table}


\subsection{LLM-as-Judge Methods}

\noindent\textbf{Foundation Models.}
To comprehensively evaluate LLM-as-judge methods across different model scales and architectures, we select a diverse set of foundation models:
\begin{itemize}
    \item Closed-source models: GPT-3.5-turbo and GPT-4o.
    \item Large-scale open-source models: DeepSeek-v3-671B and DeepSeek-r1-671B. 
    \item Medium-scale open-source models:Llama3-8B, Qwen2.5-7B and DeepSeek-r1-distill-7B. 
    \item Small-scale open-source models: Llama3-1.5B, Qwen2.5-1.5B and DeepSeek-r1-distill-1.5B. 
\end{itemize}

The DeepSeek series models are classified as \emph{reasoning models} because of their powerful reasoning capabilities, and the rest of the models are classified as \emph{general models}.
We limited our study to these models as they provide sufficient representativeness across different architectures, capabilities, and parameter scales.

This selection enables us to analyze how model size affects the performance of LLM-as-judge methods and investigate whether smaller, more computationally efficient models can achieve comparable evaluation quality to their larger counterparts.

\noindent\textbf{Existing Prompting Methods.}
We evaluate four representative prompting methods that represent different perspectives to eliciting code evaluation capabilities from LLMs:

\begin{itemize}
    \item \textbf{Vanilla}, which is a straightforward prompting method that directly asks the model to evaluate code correctness based on the problem description and implementation, without additional guidance.
    
    \item \textbf{CoT}~\cite{wei2022chain}, which encourages the model to perform step-by-step reasoning by analyzing the code's logic, identifying potential issues, and then making a final judgment on correctness.
    
    \item \textbf{ICE\_SCORE}~\cite{zhuo2024ice}, which performs multi-dimensional evaluation and instructs the LLM to predict an evaluation score from 0 to 4 based on an evaluation criterion. In our experiments, we adopt the evaluation score as 0 or 1 for functional correctness. 

    \item \textbf{CodeJudge}~\cite{tong2024codejudge}, which is a two-phase method, where a summary of the given code is first generated and then is evaluated to determine whether the code is correct, based on the generated summary and the given problem description.
\end{itemize}


\subsection{Evaluation Metrics}

To comprehensively assess the performance of LLM-as-judge methods for code evaluation, we employ three 
metrics.

\noindent\textbf{Accuracy (Acc).} It measures the proportion of correctly classified instances among all evaluated samples. For $n$ code samples with ground truth labels $y_i$ and predicted labels $\hat{y}_i$:
\begin{equation*}
\text{Accuracy} = \frac{1}{n} \sum_{i=1}^{n} \mathbb{I}(\hat{y}_i = y_i)
\end{equation*}
where $\mathbb{I}(\cdot)$ is the indicator function that returns 1 for correct predictions and 0 otherwise.

\noindent\textbf{F1 Score (F1).} It is the macro-average of precision and recall, particularly valuable for our datasets with class imbalance:
\begin{gather*}
\text{Precision} = \frac{\text{TP}}{\text{TP} + \text{FP}}, \quad \text{Recall} = \frac{\text{TP}}{\text{TP} + \text{FN}}\\
\text{F1} = 2 \times \frac{\text{Precision} \times \text{Recall}}{\text{Precision} + \text{Recall}}
\end{gather*}
F1 ranges from 0 to 1, with higher values indicating better performance in identifying functionally correct code.

\noindent\textbf{Matthews Correlation Coefficient (MCC).} It provides a balanced measure by considering all confusion matrix entries:
\begin{equation*}
\text{MCC} = \frac{\text{TP} \times \text{TN} - \text{FP} \times \text{FN}}{\sqrt{(\text{TP} + \text{FP})(\text{TP} + \text{FN})(\text{TN} + \text{FP})(\text{TN} + \text{FN})}}
\end{equation*}

MCC ranges from -1 to 1, where 1 indicates perfect prediction, 0 random prediction, and -1 inverse prediction. 
It is less sensitive to class imbalance than the accuracy and F1 score.

\subsection{Implementation Details}

Across all experiments, we fix the maximum context length at 8k tokens. 
Temperature settings were tailored to model type: 0.6 for reasoning-focused models (to promote exploratory reasoning) and 0.0 for general-purpose models (to ensure deterministic outputs).

We 
interact with the following large-scale models (via their official APIs): DeepSeek-v3-67B, DeepSeek-r1-67B, GPT-3.5-turbo and GPT-4o. 
Medium/small-scale open-source models were 
from Hugging Face, with inference optimized via VLLM~\cite{kwon2023efficient} on a single RTX 4090 GPU to maximize throughput.

\begin{table*}[htbp]
	\centering
	\caption{Performance Comparison of Different Models and Prompting Methods across Datasets}
	\rowcolors{4}{gray!15}{}
		\begin{tabular}{llcccccccccccc}
			\toprule
			\multirow{2}{*}{Base Model} & \multirow{2}{*}{Prompt} & \multicolumn{3}{c}{HumanEval-Judge} & \multicolumn{3}{c}{MBPP-Judge} & \multicolumn{3}{c}{BigCodeBench-Judge} & \multicolumn{3}{c}{Avg.} \\
			\cmidrule(lr){3-5} \cmidrule(lr){6-8} \cmidrule(lr){9-11} \cmidrule(lr){12-14}
			& & Acc & F1 & MCC & Acc & F1 & MCC & Acc & F1 & MCC & Acc & F1 & MCC \\
			\midrule
			& Vanilla & 0.730 & 0.658 & 0.319 & 0.663 & 0.642 & 0.293 & \textbf{0.584} & \textbf{0.584} & 0.219 & 0.659 & \textbf{\underline{0.628}} & 0.277\\
			GPT-3.5-turbo & CoT & \textbf{0.781} & 0.601 & 0.303 & 0.687 & 0.558 & 0.214 & 0.493 & 0.449 & 0.210 & 0.654 & 0.536 & 0.242\\
			Close-Source & ICE\_SCORE & 0.752 & 0.571 & 0.200 & 0.695 & 0.575 & 0.242 & 0.516 & 0.487 & 0.229 & 0.654 & 0.544 & 0.224 \\
			& CodeJudge & 0.773  & \textbf{0.666} & \textbf{0.343} & \textbf{0.726} & \textbf{0.661} & \textbf{0.349} & 0.525 & 0.498 & \textbf{0.246} & \textbf{\underline{0.675}} & 0.608 & \textbf{\underline{0.313}}\\
			\midrule
			& Vanilla & 0.731 & 0.679 & 0.375 & 0.645 & 0.633 & 0.293 & 0.639 & 0.614 & \textbf{0.361} & 0.672 & 0.642 & 0.343\\
			GPT-4o & CoT & 0.873 & 0.816 & 0.643 & \textbf{0.807} & \textbf{0.763} & \textbf{0.554} & 0.706 & \textbf{0.706} & 0.294 & \textbf{\underline{0.795}} & \textbf{\underline{0.762}} & 0.497\\
			Close-Source & ICE\_SCORE & \textbf{0.880} & 0.821 & 0.660 & 0.785 & 0.732 & 0.499 & 0.666 & 0.666 & 0.334 & 0.777 & 0.740 & \textbf{\underline{0.498}}\\
			& CodeJudge & \textbf{0.880} & \textbf{0.843} & \textbf{0.686} & 0.735 & 0.718 & 0.446 & \textbf{0.714} & 0.705 & 0.286 & 0.776 & 0.755 & 0.473\\
			\midrule
			& Vanilla & 0.873 & 0.822 & 0.649 & \textbf{0.792} & \textbf{0.742} & \textbf{0.516} & 0.626 & 0.623 & 0.367 & 0.764 & 0.729 & 0.511\\
			DS-v3 & CoT & 0.867 & 0.788 & 0.623 & 0.786 & 0.722 & 0.507 & 0.638 & 0.632 & 0.410 & 0.764 & 0.714 & 0.513 \\
			671B & ICE\_SCORE & 0.814 & 0.702 & 0.445 & 0.737 & 0.640 & 0.371 & 0.564 & 0.553 & 0.271 & 0.705 & 0.632 & 0.362\\
			& CodeJudge & \textbf{0.884} & \textbf{0.831} & \textbf{0.675} & 0.783 & 0.730 & 0.494 & \textbf{0.675} & \textbf{0.674} & \textbf{0.325} & \textbf{\underline{0.781}} & \textbf{\underline{0.745}} & \textbf{\underline{0.498}} \\
			\midrule
			& Vanilla & \textbf{0.925} & \textbf{0.904} & \textbf{0.812} & \textbf{0.828} & \textbf{0.806} & \textbf{0.613} & \textbf{0.748} & \textbf{0.735} & \textbf{0.471} & \textbf{\underline{0.834}} & \textbf{\underline{0.815}} & \textbf{\underline{0.632}}\\
			DS-r1 & CoT & 0.920 & 0.899 & 0.802 & 0.821 & 0.798 & 0.597 & 0.728 & 0.710 & 0.423 & 0.823 & 0.802 & 0.607 \\
			671B & ICE\_SCORE & \textbf{0.925} & \textbf{0.904} & 0.811 & 0.825 & 0.801 & 0.604 & 0.744 & 0.730 & 0.461 & 0.831 & 0.812 & 0.625\\
			& CodeJudge & 0.897 & 0.875 & 0.765 & 0.791 & 0.773 & 0.547 & 0.731 & 0.708 & 0.428 & 0.806 & 0.785 & 0.580\\
			\midrule
			\midrule
			& Vanilla & 0.639 & 0.581 & 0.184 & 0.645 & 0.622 & 0.252 & 0.581 & 0.560 & 0.147 & 0.622 & 0.588 & 0.194 \\
			Llama3 & CoT & 0.667 & 0.608 & 0.235 & 0.682 & \textbf{0.657} & \textbf{0.318} & 0.569 & \textbf{0.569} & \textbf{0.178} & 0.639 & 0.611 & 0.244\\
			8B & ICE\_SCORE & \textbf{0.738} & \textbf{0.660} & \textbf{0.320} & \textbf{0.700} & 0.656 & 0.315 & 0.536 & 0.533 & 0.161 & \textbf{\underline{0.658}} & \textbf{\underline{0.616}} & \textbf{\underline{0.265}}\\
			& CodeJudge & 0.509 & 0.496 & 0.121 & 0.562 & 0.559 & 0.184 & \textbf{0.585} & 0.563 & 0.127 & 0.552 & 0.539 & 0.144\\
			\midrule
			& Vanilla & 0.769 & 0.662 & 0.333 & \textbf{0.748} & \textbf{0.694} & \textbf{0.408} & 0.575 & 0.563 & 0.305 & 0.697 & 0.640 & 0.349 \\
			Qwen2.5 & CoT & \textbf{0.789} & 0.673 & 0.372 & 0.743 & 0.684 & 0.394 & 0.563 & 0.545 & 0.306 & 0.698 & 0.634 & 0.357\\
			7B & ICE\_SCORE & \textbf{0.789} & 0.684 & 0.384 & 0.745 & 0.684 & 0.398 & \textbf{0.591} & 0.582 & \textbf{0.327} & \textbf{\underline{0.708}} & 0.650 & \textbf{\underline{0.370}} \\
			& CodeJudge & 0.783 & \textbf{0.694} & \textbf{0.391} & 0.739 & 0.693 & 0.395 & 0.589 & \textbf{0.584} & 0.287 & 0.704 & \textbf{\underline{0.657}} & 0.358\\
			\midrule
			& Vanilla & 0.816 & \textbf{0.770} & \textbf{0.546} & \textbf{0.766} & \textbf{0.730} & \textbf{0.464} & \textbf{0.629} & \textbf{0.629} & \textbf{0.319} & \textbf{\underline{0.737}} & \textbf{\underline{0.710}} & \textbf{\underline{0.443}} \\
			DS-r1-distill & CoT & 0.773 & 0.705 & 0.410 & 0.765 & 0.723 & 0.456 & 0.524 & 0.507 & 0.197 & 0.687 & 0.645 & 0.354 \\
			7B & ICE\_SCORE & \textbf{0.817} & 0.753 & 0.506 & 0.747 & 0.698 & 0.410 & 0.571 & 0.565 & 0.262 & 0.712 & 0.672 & 0.393\\
			& CodeJudge & 0.788 & 0.720 & 0.440 & 0.718 & 0.676 & 0.355 & 0.604 & 0.604 & 0.265 & 0.703 & 0.667 & 0.353\\
			\midrule
			\midrule
			& Vanilla & 0.267 & 0.235 & -0.030 & 0.343 & 0.277 & -0.051 & 0.551 & 0.405 & -0.092 & 0.387 & 0.306 & -0.058\\
			Llama3 & CoT & 0.603 & 0.475 & -0.050 & 0.553 & 0.490 & -0.019 & 0.540 & \textbf{0.540} & \textbf{0.125} & 0.565 & \textbf{\underline{0.502}} & 0.019 \\
			1B & ICE\_SCORE & \textbf{0.625} & \textbf{0.523} & \textbf{0.049} & \textbf{0.610} & \textbf{0.496} & \textbf{0.025} & 0.479 & 0.479 & 0.000 & \textbf{\underline{0.571}} & 0.499 & \textbf{\underline{0.025}} \\
			& CodeJudge & 0.400 & 0.399 & 0.000 & 0.401 & 0.400 & -0.082 & \textbf{0.561} & 0.487 & 0.013 & 0.454 & 0.429 & -0.023 \\
			\midrule
			& Vanilla & 0.630 & \textbf{0.567} & \textbf{0.155} & \textbf{0.663} & \textbf{0.630} & \textbf{0.262} & \textbf{0.586} & \textbf{0.573} & \textbf{0.147} & 0.626 & \textbf{\underline{0.590}} & \textbf{\underline{0.188}} \\
			Qwen2.5 & CoT & 0.684 & 0.533 & 0.075 & 0.650 & 0.546 & 0.133 & 0.439 & 0.376 & 0.076 & 0.591 & 0.485 & 0.095\\
			1.5B & ICE\_SCORE & 0.686 & 0.480 & -0.014 & 0.651 & 0.522 & 0.111 & 0.446 & 0.394 & 0.078 & 0.594 & 0.465 & 0.058\\
			& CodeJudge & \textbf{0.717} & 0.512 & 0.069 & 0.658 & 0.522 & 0.123 & 0.561 & 0.487 & 0.095 & \textbf{\underline{0.645}} & 0.507 & 0.096\\
			\midrule
			& Vanilla & 0.728 & 0.639 & 0.278 & \textbf{0.714} & \textbf{0.664} & \textbf{0.336} & 0.514 & 0.510 & 0.109 & \textbf{\underline{0.652}} & \textbf{\underline{0.604}} & \textbf{\underline{0.241}} \\
			DS-r1-distill & CoT & 0.728 & 0.575 & 0.172 & 0.702 & 0.612 & 0.275 & 0.466 & 0.435 & 0.087 & 0.632 & 0.541 & 0.178\\
			1.5B & ICE\_SCORE & \textbf{0.747} & \textbf{0.658} & \textbf{0.317} & 0.702 & 0.643 & 0.299 & 0.480 & 0.467 & 0.072 & 0.643 & 0.589 & 0.229\\
			& CodeJudge & 0.713 & 0.570 & 0.152 & 0.648 & 0.547 & 0.131 & \textbf{0.549} & \textbf{0.549} & \textbf{0.152} & 0.637 & 0.555 & 0.145\\
			\bottomrule
		\end{tabular}
	\label{tab:model_performance}
\end{table*}

\subsection{Empirical Findings}
The results are shown in Table \ref{tab:model_performance}.
Based on the  extensive experiments with different models and prompting methods for code evaluation tasks, we have identified differences between general models (GPT/DeepSeek-V3/Llama3/Qwen2.5 series) and reasoning models (DeepSeek-R1 series):

\smallskip
\noindent\textbf{(1) General Models Depend on Prompt Engineering.}
Our analysis reveals that general-purpose models show high sensitivity to prompt engineering.

Large-scale models respond differently to prompts: GPT-3.5-turbo and DeepSeek-v3 perform best with CodeJudge, while GPT-4o excels with CoT. 
For medium and small-scale models, structured approaches like ICE\_SCORE significantly improve performance. 

A notable example is Llama3 8B, which achieved an accuracy of 0.658 and MCC of 0.265 using ICE\_SCORE, substantially outperforming its Vanilla baseline (accuracy 0.622, MCC 0.194).

\begin{tcolorbox}[colback=SeaGreen!10!CornflowerBlue!10,colframe=RoyalPurple!55!Aquamarine!100!,title=Finding 1]
For general models, optimal prompting strategies vary by architecture and scale, requiring model-specific customization.
\end{tcolorbox}

\noindent\textbf{(2) Reasoning Models Prefer Simple Prompts.}
In contrast to their general counterparts, reasoning models exhibit consistent superior performance with simpler prompts. 
The Vanilla method emerges as the most effective approach across all DeepSeek-r1-distill model sizes (7B and 1.5B). 

Notably, increased prompt complexity often leads to performance degradation, with the 7B model achieving a remarkable 0.737 accuracy using the basic Vanilla approach.

\begin{tcolorbox}[colback=SeaGreen!10!CornflowerBlue!10,colframe=RoyalPurple!55!Aquamarine!100!,title=Finding 2]
    For reasoning models, they have already internalized the reasoning capabilities, requiring no external provision of reasoning steps or structured frameworks. 
\end{tcolorbox}

\begin{figure*}[h]
    \centering
    \includegraphics[width=1\textwidth]{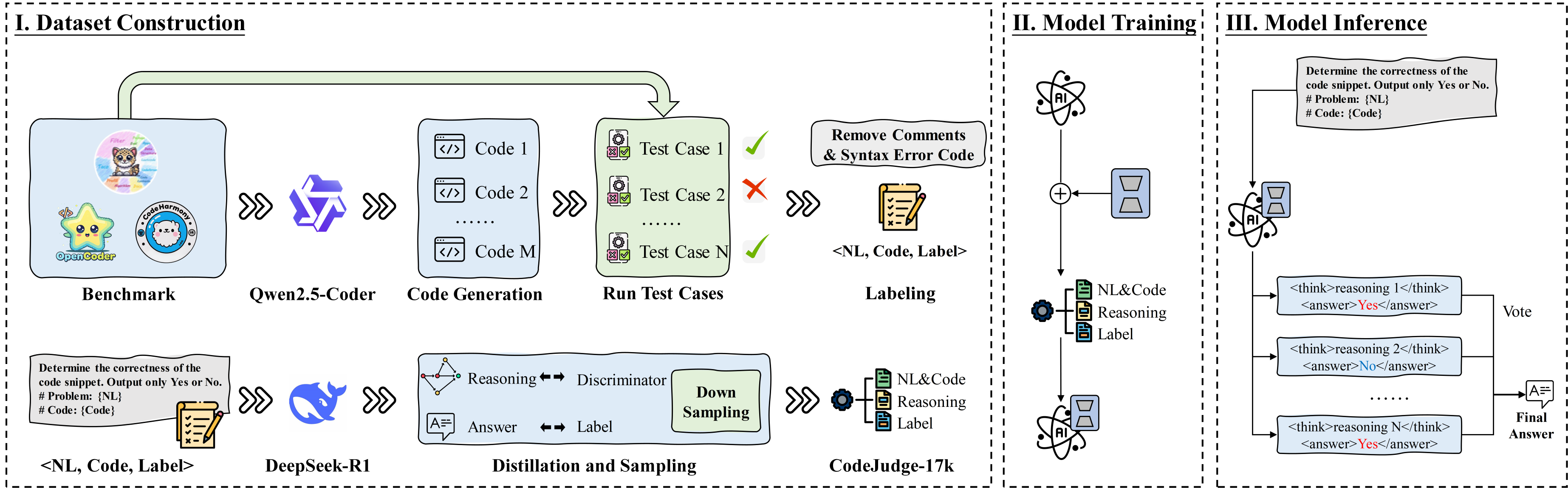}
    \caption{The overall method of {\tool}.}
    \label{fig:method}
\end{figure*}

\noindent\textbf{(3) Performance Comparison.}
Our evaluation reveals the superior stability of reasoning models across diverse datasets, highlighting their robustness and generalizability.

For large-scale models, DeepSeek-r1-671B with the best prompting method achieves an accuracy of 0.834, F1 score of 0.815 and MCC of 0.632, significantly higher than others.
Similarly, for the 7B-scale, DeepSeek-r1-distill 7B with the best prompting method achieves an accuracy of 0.737, F1 score of 0.710, and MCC of 0.443.
For the 1.5B-scale models, DeepSeek-r1-distill 1.5B achieves the best accuracy of 0.652, F1 score of 0.604 and MCC of 0.241.

\begin{tcolorbox}[colback=SeaGreen!10!CornflowerBlue!10,colframe=RoyalPurple!55!Aquamarine!100!,title=Finding 3]
    At comparable parameter scales, reasoning models demonstrate superior and more stable performance across different datasets compared to general models. 
\end{tcolorbox}

\section{Methods}
\label{sec:method}

In this section, we introduce our method {\tool}. Based on the empirical findings in Section~\ref{sec:empirical}, we build on two key insights: 
(1) explicit reasoning paths significantly enhance code evaluation accuracy while enabling better sample explainability;
and 
(2) smaller models with appropriate training can potentially match or exceed the performance of much larger models. 

{\tool} distills reasoning capabilities into compact models to balance accuracy with computational efficiency, as shown in Figure~\ref{fig:method}.

\subsection{Dataset Construction}

To effectively transfer reasoning capabilities from large-scale models to {\tool}, high-quality training data are essential. 

\subsubsection{Source Benchmark Collection} 

We 
follow three key principles for dataset selection: 
\begin{itemize}
    \item \textbf{Diversity:} The dataset needs to cover a wide range of programming scenarios, including algorithmic problems, system programming, and library usage.
    \item \textbf{Difficulty:} In addition to basic syntax tasks, the dataset must encompass complex logical challenges and multi-step reasoning problems.
    \item \textbf{Quality:} The dataset needs to contain high-coverage test cases to ensure reliable functional correctness assessment. 
\end{itemize}

Based on these principles, we 
select three large-scale code generation benchmarks, i.e, KodCode~\cite{xu2025kodcode}, OpenCoder~\cite{huang2024opencoder}, and CodeHarmony~\cite{yang2024less}, as the seed data.
These benchmarks are widely used to train large-scale CodeLLMs. 

\subsubsection{Code Generation and Labeling} 
To ensure a balanced and representative training set, we implement a systematic data generation and labeling process. 
In addition, to control the distribution of correct and incorrect examples in our dataset, we generate multiple candidate solutions using Qwen2.5-Coder (1.5B/7B) for each programming task. For quality assurance, we employ a multi-step validation process as HumanEval-Judge:
\begin{itemize}
    \item We evaluate each solution's functional correctness through test cases and compute the pass@1 as the label.
    \item We apply static analysis tools to identify and filter out solutions containing syntax errors.
    \item We remove code comments to focus the evaluation on core implementation logic.
\end{itemize}


\subsubsection{Reasoning Knowledge Distillation} 
To transfer the logical reasoning capabilities of large-scale reasoning models to our target dataset and enhance sample explainability, we implement a distillation process. 
For each triple $\langle$nl, code, label$\rangle$, we use DeepSeek-R1-671B (the SOTA reasoning model, as shown in Table~\ref{tab:model_performance}) in Vanilla setting to produce independent judgments on code functional correctness, including both predicted labels and reasoning paths. 
This process yields raw distillation data in the format $\langle$nl, code, label, reasoning$\rangle$.

\subsubsection{Data Filtering and Sampling} 
We implement a multi-stage filtering mechanism: 
\begin{itemize}
    \item \textbf{Accuracy filtering.} We remove samples where DeepSeek-R1-671B's predictions disagreed with test case labels to ensure consistency; 
    \item \textbf{Logical coherence filtering.} We employ DeepSeek-V3 as a discriminator\footnote{The prompt can be found in the our GitHub repository.} to detect and eliminate reasoning paths containing hallucinations or logical inconsistencies; 
    \item \textbf{Class balancing.} We downsample the filtered data to achieve a 1:1 ratio between positive and negative samples, addressing the imbalance in the original dataset where correct samples were overrepresented.
\end{itemize}

As a result, we construct \textsc{CodeJudge-17k}, a high-quality dataset containing 17,000 samples. 
\textsc{CodeJudge-17k} features a balanced distribution of correct and incorrect code samples across diverse programming tasks, spanning from basic algorithmic challenges to complex system implementations. 
Each sample is accompanied by a detailed reasoning path that explains the judgment process, making the dataset valuable for training explainable code judgment models. 

\subsection{Model Training}

To transfer reasoning capabilities to smaller models while maintaining efficiency, we train the model in three stages. 

\noindent \textbf{1. Knowledge Injection}
We hypothesize that explicit reasoning paths are crucial for code evaluation tasks. 
To inject this capability while minimizing deployment costs, we use DS-R1-distil (1.5B/7B) as base models which are fine-tuned on \textsc{CodeJudge-17k}. 
This enables smaller models to learn from larger experts while requiring only 1\% of the parameters.

\noindent \textbf{2. Parameter-Efficient Fine-tuning with LoRA}
To optimize model training while maintaining performance, we adopt Low-Rank Adaptation (LoRA), a parameter-efficient fine-tuning technique. 
This freezes pre-trained weights $W_0 \in \mathbb{R}^{d \times k}$ and introduces trainable low-rank matrices $A \in \mathbb{R}^{r \times k}$, $B \in \mathbb{R}^{d \times r}$ ($r \ll \min(d,k)$):
\begin{equation*}
W = W_0 + BA
\end{equation*}
This reduces trainable parameters from $dk$ to $r(d+k)$, preserving performance with minimal overhead.

\noindent \textbf{3. PiSSA Initialization}
To enhance training efficiency and model performance, we leverage Principal Singular Vector Adaptation (PiSSA)~\cite{meng2024pissa} for initializing LoRA matrices. 
Instead of Kaiming-uniform~\cite{he2015delving} initialization used in LoRA, PiSSA leverages the intrinsic low-rank structure of $W_0$ through truncated SVD, i.e., 
%
    $W_0 \approx U_r \Sigma_r V_r^\top$.  
%
The LoRA matrices are then initialized as 
\begin{equation*}
    B = U_r \Sigma_r^{1/2}, \quad A = \Sigma_r^{1/2} V_r^\top
\end{equation*}
This ensures $\Delta W = BA$ initially aligns with $W_0$'s principal subspace, concentrating updates on directions critical for functional preservation. 
Compared to the Kaiming-uniform initialization, PiSSA provides structured starting points that improve convergence speed and final performance, particularly in low-rank regimes.

\subsection{Model Inference}
Considering that the reasoning model may have inconsistent reasoning paths when the temperature is set to 0.6, we use the Majority Vote strategy to determine the final reasoning result and further enhance model inference performance.
This belongs to parallel inference methods, where the model performs multiple independent inferences on the same input, and the most frequent result is selected as the final judgment.

From a probabilistic perspective, if the probability of a correct judgment in a single inference is $P(A)$,  the probability of the final result being correct can be modeled through a binomial distribution when conducting $T$ independent inferences. 
Specifically, if at least $(T+1)/2$ inference results are correct (i.e., the majority vote is correct), then the probability of the final judgment being correct is 
\begin{equation*}
P\left(X \geq \frac{T+1}{2}\right) = \sum_{k=\lceil\frac{T+1}{2}\rceil}^{T} \binom{T}{k} P(A)^k (1-P(A))^{T-k}.
\end{equation*}
%
%
When $P(A) > 0.5$, according to the Law of Large Numbers, as $T$ increases, the success probability of the majority vote strategy $P\left(X \geq \frac{T+1}{2}\right)$ will continuously improve. 
This explains why majority voting can effectively enhance model performance: as long as the accuracy of a single inference exceeds random guessing (i.e., $P(A) > 0.5$), multiple voting can significantly reduce the probability of misjudgment.

In our experiments, we perform $T=7$ independent inferences for each test sample and use majority voting to determine the final judgment result. Note that $T$ is set $7$ based on RQ3 findings (Section~\ref{sect:rq3}) as the optimal trade-off between model performance and inference latency.

\section{Experiments and Analysis}
\label{sec:experiments}

To evaluate the effectiveness and benefits of {\tool}, we mainly study the following three research questions (RQs):

\subsection{RQ1: How does {\tool} perform compared to the state-of-the-art methods?}

\begin{table*}[htbp]
\centering
\caption{Performance Comparison of Different Models and Prompting Methods across Datasets}
\rowcolors{4}{gray!15}{}
\begin{tabular}{llcccccccccccc}
    \toprule
    \multirow{2}{*}{Base Model} & \multirow{2}{*}{Prompt} & \multicolumn{3}{c}{HumanEval-Judge} & \multicolumn{3}{c}{MBPP-Judge} & \multicolumn{3}{c}{BigCodeBench-Judge} & \multicolumn{3}{c}{Avg.} \\
    \cmidrule(lr){3-5} \cmidrule(lr){6-8} \cmidrule(lr){9-11} \cmidrule(lr){12-14}
    & & Acc & F1 & MCC & Acc & F1 & MCC & Acc & F1 & MCC & Acc & F1 & MCC \\
    \midrule
    GPT-3.5-turbo & CodeJudge & 0.773  & 0.666 & 0.343 & 0.726 & 0.661 & 0.349 & 0.525 & 0.498 & 0.246 & 0.675 & 0.608 & 0.313\\
    GPT-4o & CoT & 0.873 & 0.816 & 0.643 & 0.807 & 0.763 & 0.554 & 0.706 & 0.706 & 0.294 & 0.795 & 0.762 & 0.497\\
    DS-v3 671B & CodeJudge & 0.884 & 0.831 & 0.675 & 0.783 & 0.730 & 0.494 & 0.675 & 0.674 & 0.325 & 0.781 & 0.745 & 0.498 \\
    DS-r1 671B & Vanilla & 0.925 & 0.904 & 0.812 & 0.828 & 0.806 & 0.613 & 0.748 & 0.735 & 0.471 & 0.834 & 0.815 & 0.632\\
    \midrule
    Llama3 8B & ICE\_SCORE & 0.738 & 0.660 & 0.320 & 0.700 & 0.656 & 0.315 & 0.536 & 0.533 & 0.161 & 0.658 & 0.616 & 0.265\\
    Qwen2.5 7B & ICE\_SCORE & 0.789 & 0.684 & 0.384 & 0.745 & 0.684 & 0.398 & 0.591 & 0.582 & 0.327 & 0.708 & 0.650 & 0.370 \\
    DS-r1-distill 7B & Vanilla & 0.816 & 0.770 & 0.546 & 0.766 & 0.730 & 0.464 & 0.629 & 0.629 & 0.319 & 0.737 & 0.710 & 0.443 \\
    {\tool} 7B & Vanilla & \textbf{0.883} & \textbf{0.847} & \textbf{0.695} & \textbf{0.806} & \textbf{0.782} & \textbf{0.564} & \textbf{0.729} & \textbf{0.717} & \textbf{0.435} & \textbf{\underline{0.806}} & \textbf{\underline{0.782}} & \textbf{\underline{0.565}}\\
    \midrule
    Llama3 1B & ICE\_SCORE & 0.625 & 0.523 & 0.049 & 0.610 & 0.496 & 0.025 & 0.479 & 0.479 & 0.000 & 0.571 & 0.499 & 0.025 \\
    Qwen2.5 1.5B & Vanilla & 0.630 & 0.567 & 0.155 & 0.663 & 0.630 & 0.262 & 0.586 & 0.573 & 0.147 & 0.626 & 0.590 & 0.188 \\
    DS-r1-distill 1.5B & Vanilla & 0.728 & 0.639 & 0.278 & 0.714 & 0.664 & 0.336 & 0.514 & 0.510 & 0.109 & 0.652 & 0.604 & 0.241 \\
    {\tool} 1.5B & Vanilla & \textbf{0.842} & \textbf{0.799} & \textbf{0.601} & \textbf{0.778} & \textbf{0.755} & \textbf{0.510} & \textbf{0.681} & \textbf{0.653} & \textbf{0.318} & \textbf{\underline{0.767}} & \textbf{\underline{0.736}} & \textbf{\underline{0.476}}\\
    \bottomrule
\end{tabular}
\label{tab:rq1}
\end{table*}



To evaluate the performance of {\tool}, we compare it with various models mentioned in Section~\ref{sec:empirical}. For a fair comparison, we use the most effective prompt for each model and employ the same evaluation metrics. The results are presented in Table~\ref{tab:rq1}.


\noindent\textbf{(1) Performance Comparison.}
Both {\tool} 1.5B and 7B models significantly outperform other models in their respective parameter scales, with substantial improvements across accuracy, F1 score and MCC metrics. In particular,  
{\tool} 1.5B surpasses Llama3 1B, Qwen2.5 1.5B and even the base DS-r1-distill 1.5B model by large margins. 
Similarly, {\tool} 7B shows clear advantages over Llama3 8B, Qwen2.5 7B, and the base DS-r1-distill 7B model. 

\noindent\textbf{(2) Parameter Efficiency.}
The parameter efficiency of our method is particularly noteworthy, as {\tool} 1.5B achieves performance comparable to DS-r1-distill 7B despite using only about 20\% of its parameters, demonstrating the effectiveness of our knowledge distillation method in transferring reasoning capabilities to smaller models. 

Most impressively, {\tool} 7B outperforms both (closed-source) GPT-4o and DeepSeek-V3 (671B) across all three datasets, falling short only of DeepSeek-R1 671B. 
This is remarkable considering that {\tool} 7B uses only about 1\% of the parameters of these larger models.

Both {\tool} variants maintain strong performance across all evaluation datasets, indicating robust generalization capabilities. 
These results validate our hypothesis that explicit reasoning paths are crucial for code evaluation tasks and demonstrate that smaller models can effectively learn these reasoning patterns through our proposed fine-tuning method.

\begin{tcolorbox}
[colback=SeaGreen!10!CornflowerBlue!10,colframe=RoyalPurple!55!Aquamarine!100!,title=Summary of RQ1]
{\tool} demonstrates superior performance in code evaluation compared to state-of-the-art methods. 
The 1.5B variant outperforms all models in its parameter class, matching 
models of 5x larger. 
The 7B variant surpasses GPT-4o and DeepSeek-V3(671B), using only 1\% of their parameters.
\end{tcolorbox}

\begin{figure}[htbp]
    \centering
    \begin{subfigure}[b]{0.24\textwidth}
        \centering
        \includegraphics[width=\textwidth]{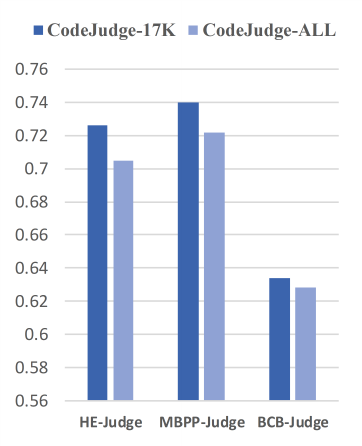}
        \caption{{\tool} 1.5B}
        \label{fig:data_ablation_1}
    \end{subfigure}
    \hfill
    \begin{subfigure}[b]{0.24\textwidth}
        \centering
        \includegraphics[width=\textwidth]{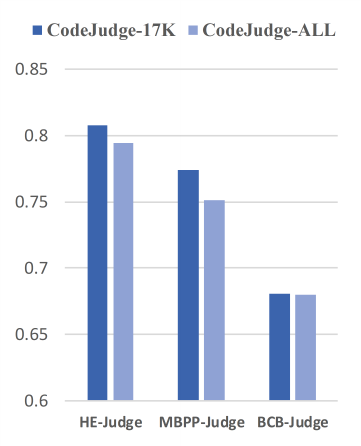}
        \caption{{\tool} 7B}
        \label{fig:data_ablation_2}
    \end{subfigure}
    \caption{Ablation Study (F1 Score) of Data Filtering Component}
    \label{fig:data_ablation}
\end{figure}

\subsection{RQ2: What is the impact of different components of {\tool}?}

To evaluate the effectiveness of different components of {\tool}, we conducted a series of ablation studies, focusing on three key aspects: 
data filtering, parameter initialization and inference strategy.

\noindent\textbf{(1) Data Filtering Component.}
Figure \ref{fig:data_ablation} illustrates the impact of the data filtering component on model performance. 
We compare the F1 scores under k=1 (single inference) across different datasets and observe that the data filtering strategy consistently and significantly improves model performance. 
This empirical evidence strongly supports the hypothesis that high-quality reasoning paths are crucial for models to develop accurate code evaluation capabilities.

Specifically, the relative improvement from filtering is notably more pronounced in the smaller 1.5B model compared to the 7B model. This distinct impact suggests that smaller models, with their inherently limited representational capacity, benefit disproportionately from high-quality training data, as they lack the parameter space to effectively learn from noisy or ambiguous examples. 

\begin{figure}[htbp]
    \centering
    \begin{subfigure}[b]{0.24\textwidth}
        \centering
        \includegraphics[width=\textwidth]{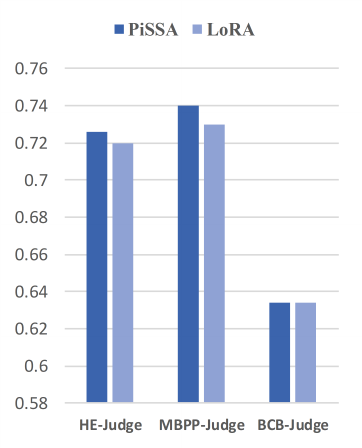}
        \caption{{\tool} 1.5B}
        \label{fig:pissa_ablation_1}
    \end{subfigure}
    \hfill
    \begin{subfigure}[b]{0.24\textwidth}
        \centering
        \includegraphics[width=\textwidth]{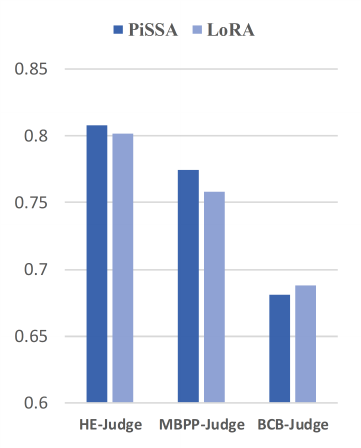}
        \caption{{\tool} 7B}
        \label{fig:pissa_ablation_2}
    \end{subfigure}
    \caption{Ablation Study (F1 Score) of PiSSA Component}
    \label{fig:pissa_ablation}
\end{figure}

\noindent\textbf{(2) PiSSA Component.}
Figure \ref{fig:pissa_ablation} shows the impact of PiSSA initialization on model performance. 
We also compare F1 scores at k=1 across different initialization methods to isolate this component's contribution. 
In standard LoRA implementations, the A matrix is typically initialized using Kaiming-uniform initialization, while the B matrix is initialized to zero. 
In contrast, PiSSA derives both A and B matrices through SVD decomposition, which fundamentally aligns the initialization with model's intrinsic parameter structure.

The experimental results reveal that PiSSA yields substantial performance improvements on the HumanEval-Judge and MBPP-Judge datasets compared to standard LoRA initialization techniques. 
However, we observe that the performance enhancement on the more challenging BigCodeBench-Judge dataset is less pronounced, suggesting that initialization benefits may vary with task complexity and dataset characteristics.

These findings indicate that PiSSA initialization helps models converge to more optimal solution spaces, particularly in parameter-constrained low-rank adaptation scenarios. 

\begin{figure}[htbp]
    \centering
    \begin{subfigure}[b]{0.36\textwidth}
        \centering
        \includegraphics[width=\textwidth]{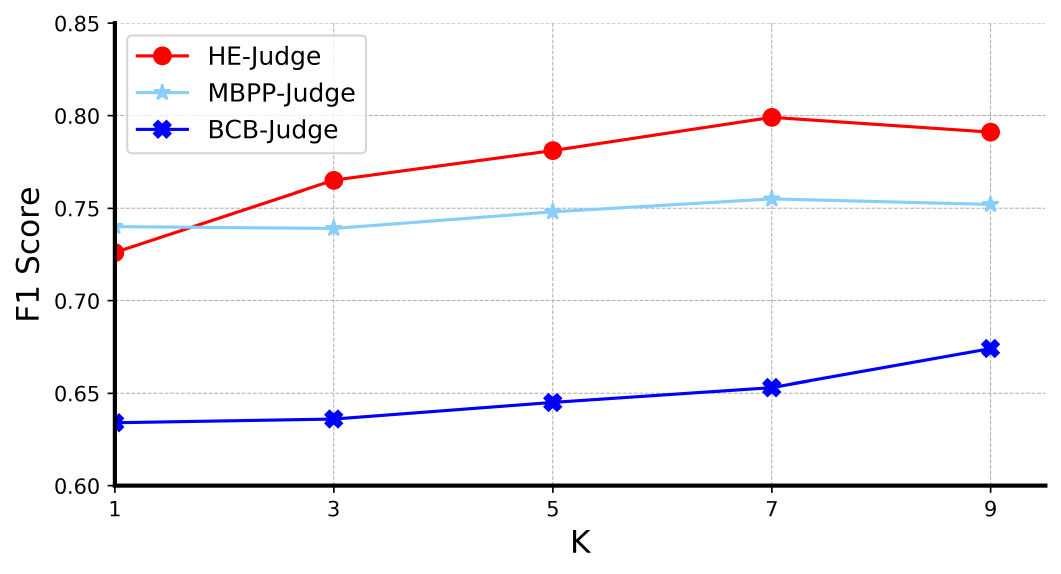}
        \caption{{\tool} 1.5B}
        \label{fig:K_ablation_1}
    \end{subfigure}
    \hfill
    \begin{subfigure}[b]{0.36\textwidth}
        \centering
        \includegraphics[width=\textwidth]{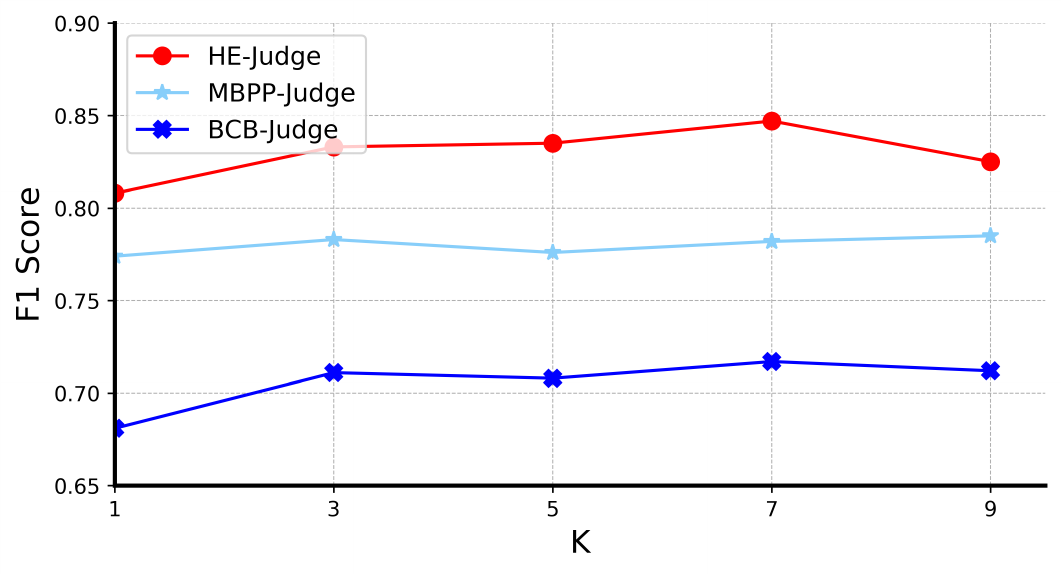}
        \caption{{\tool} 7B}
        \label{fig:K_ablation_2}
    \end{subfigure}
    \caption{Ablation study (F1 Score) of the inference component}
    \label{fig:K_ablation}
\end{figure}

\noindent\textbf{(3) Inference Component.}
Figure \ref{fig:K_ablation} presents a detailed analysis of how our inference strategy affects model performance. 
We systematically compare F1 scores across different values of k (the number of inference passes) to identify the optimal configuration. 
The results demonstrate a clear pattern: as k increases, model performance consistently improves, though with diminishing returns at higher values. 

To determine the most practical configuration for real-world applications, we conduct an 
analysis of the performance-efficiency trade-off. 
Our experiments are performed using vllm as the inference server on a single NVIDIA RTX 4090 GPU. 
The baseline latency (k=1) for a single inference pass is 0.15s and 0.30s for the 1.5B and 7B models, respectively. 
As expected, the time cost scales linearly with k, reaching approximately 1s (1.5B) and 2s (7B) when k=7.

By analyzing both the performance improvements and computational overhead across different k values, we identify k=7 as the optimal one. 
This configuration delivers substantial accuracy gains while maintaining reasonable inference latency, making it well-suited for practical applications where both prediction quality and response time are critical considerations.

\begin{tcolorbox}[colback=SeaGreen!10!CornflowerBlue!10,colframe=RoyalPurple!55!Aquamarine!100!,title=Summary of RQ2]
    Our ablation studies demonstrate that each component of {\tool} contributes significantly to its overall performance. 
    With the combination of data filtering, PiSSA initialization, and the optimal inference strategy, {\tool} achieves state-of-the-art performance while maintaining computational efficiency.
\end{tcolorbox}

\subsection{RQ3: Does {\tool} suffer from preference leakage?} \label{sect:rq3}

Preference leakage~\cite{li2025preference} refers to a 
contamination issue in LLM-as-judge frameworks where correlations between the synthetic data generator and the LLM-based evaluator lead to biased assessments.  

In our training process, we have used code generated by models in the same families (DeepSeek and Qwen Coder) that serve as our base models. This raises a legitimate concern: does {\tool} exhibit preference bias toward code generated by models similar to those used in its training data? 

To systematically investigate this potential issue, we consider Agreement Rate and Cohen's Kappa~\cite{cohen1960coefficient} as the evaluation metrics.
Specifically, Agreement Rate measures the consistency of judgments between different evaluation scenarios:
\begin{equation*}
    \text{Agreement Rate} = \frac{\text{Number of consistent judgments}}{\text{Total number of samples}}
\end{equation*}  
Cohen's Kappa quantifies the agreement between evaluators while accounting for chance agreement:
\begin{equation*}
    \text{Cohen's Kappa} = \frac{p_o - p_e}{1 - p_e}
\end{equation*}
where $p_o$ is the observed agreement rate and $p_e$ is the expected agreement rate by chance. 
The chance agreement $p_e$ is calculated based on the marginal distributions of each evaluator's judgments:
\begin{equation*}
    p_e = \sum_{i} (p_{i1} \times p_{i2})
\end{equation*}
where $p_{i1}$ and $p_{i2}$ represent the proportion of samples classified as category $i$ by the first and second evaluator, respectively. 
This adjustment for chance agreement makes Cohen's Kappa a more robust measure than simple agreement rate, especially when the distribution of categories is imbalanced.

We carry out experiments the assess the consistency of {\tool} from different perspectives. 

\noindent\textbf{(1) Consistency across different code generators.}
This experiment evaluates whether {\tool} maintains consistent judgments when evaluating code generated by different models for the same programming task. 
We selected 50 problems from each dataset and used two models not involved in our training data generation (i.e., GPT-4o and Claude-3.5) to generate code solutions. 
We then assessed whether {\tool} produced consistent evaluations regardless of the code's source.

\begin{table}[t]
    \centering
    \caption{Consistency analysis across different code generation models}
    \label{tab:consistency_code}
    \begin{tabular}{cccc}
    \toprule
    \textbf{Model} & \textbf{Dataset} & \textbf{Agreement Rate} & \textbf{Kappa}\\
    \midrule
    \multirow{3}{*}{GPT-4o} & HumanEval-Judge & 98.0\% & 0.96 \\
     & MBPP-Judge & 96.0\% & 0.92 \\
     & BigCodeBench-Judge & 94.0\% & 0.88 \\
    \midrule
    \multirow{3}{*}{Claude-3.5} & HumanEval-Judge & 97.0\% & 0.94 \\
     & MBPP-Judge & 95.0\% & 0.90 \\
     & BigCodeBench-Judge & 93.0\% & 0.86 \\
    \bottomrule
    \end{tabular}
\end{table}

As shown in Table \ref{tab:consistency_code}, {\tool} demonstrates high consistency in its judgments across different code generators, with agreement rates exceeding 93\% across all datasets. 
The exceptionally high Cohen's Kappa values (ranging from 0.86 to 0.96) indicate near-perfect agreement beyond what would be expected by chance. 
This 
consistency is particularly evident on the HumanEval-Judge dataset, where agreement rates reach 98\% with GPT-4o-generated code and 97\% with Claude-3.5-generated code. 
Even on the more challenging BigCodeBench-Judge dataset, which involves complex library interactions, {\tool} maintains agreement rates of 94\% and 93\% respectively. 
These results strongly suggest that {\tool}'s evaluation mechanism focuses on the intrinsic quality and correctness of code rather than superficial patterns associated with specific code generators.

\smallskip
\noindent\textbf{(2) Consistency across different problem descriptions.}
This experiment examines whether {\tool} maintains consistent judgments when the same code is evaluated against semantically equivalent--but differently phrased--problem descriptions. 
We also select 50 code samples from each dataset and use GPT-4o and Claude-3.5 to generate paraphrased versions of the original problem descriptions while preserving their semantic meaning. 
We then evaluate whether {\tool}'s judgments remained consistent across these different problem formulations.

\begin{table}[t]
    \centering
    \caption{Consistency analysis across different problem descriptions}
    \label{tab:consistency_prompt}
    \begin{tabular}{cccc}
    \toprule
    \textbf{Model} & \textbf{Dataset} & \textbf{Agreement Rate} & \textbf{Kappa}\\
    \midrule
    \multirow{3}{*}{GPT-4o} & HumanEval-Judge & 96.0\% & 0.92 \\
     & MBPP-Judge & 95.0\% & 0.90 \\
     & BigCodeBench-Judge & 94.0\% & 0.88 \\
    \midrule
    \multirow{3}{*}{Claude-3.5} & HumanEval-Judge & 95.0\% & 0.90 \\
     & MBPP-Judge & 94.5\% & 0.89 \\
     & BigCodeBench-Judge & 94.0\% & 0.87 \\
    \bottomrule
    \end{tabular}
\end{table}

Table~\ref{tab:consistency_prompt} shows that {\tool} maintains even higher consistency 
with agreement rates of 94--96\% across datasets. 
The Cohen's Kappa values (0.87-0.92) indicate near-perfect agreement, substantially exceeding what would be expected by chance. 
Notably, the consistency remains 
stable across all three datasets, with minimal variation between HumanEval-Judge, MBPP-Judge and BigCodeBench-Judge. 
The stability is particularly significant for BigCodeBench-Judge, where the complexity of library interactions could potentially make the model more sensitive to variations in problem descriptions. 
The high agreement rates for both GPT-4o and Claude-3.5 paraphrases demonstrate that {\tool} robustly captures the semantic relationship between code and requirements, focusing on functional alignment rather than superficial textual patterns in the problem description. 
This resilience to paraphrasing suggests that {\tool} has developed a deep understanding of programming tasks that transcends specific wording choices.

\begin{tcolorbox}[colback=SeaGreen!10!CornflowerBlue!10,colframe=RoyalPurple!55!Aquamarine!100!,title=Summary of RQ3]
{\tool} does not suffer from significant preference leakage. 
It maintains high consistency when evaluating code from different generators and when assessing code against semantically equivalent problem descriptions. 
\end{tcolorbox}

\section{Threats to Validity}
\label{sec:threats}

\noindent\textbf{Internal Validity.}
The primary threat to internal validity concerns implementation fidelity. 
We mitigated this by carefully implementing baseline methods according to their original descriptions, using public implementations where available, and thoroughly validating our {\tool} implementation. 
Regarding potential bias in the distilled \textsc{CodeJudge-17K} training dataset, we employed multi-stage filtering to ensure high-quality reasoning paths and accurate labels.

\smallskip
\noindent\textbf{External Validity.}
External validity threats stem from our dataset and model selections. 
We chose HumanEval-plus, MBPP-plus, and BigCodeBench for their high-quality test cases and diverse programming scenarios, though future work could explore additional programming paradigms and domain-specific languages. 
Our model selection spans various scales and architectures (closed-source GPT models, large-scale DeepSeek models, and smaller open-source models ranging from 1.5B to 8B parameters), providing meaningful insights within our hardware constraints (single RTX 4090 GPU).

\smallskip
\noindent\textbf{Construct Validity.}
Construct threats concern the performance metrics used to evaluate the performance of {\tool} and the compared methods.
To evaluate the performance of models, we utilized Accuracy, F1-score, and MCC as the evaluation metrics. 
Furthermore, to evaluate the performance leakage issue of {\tool}, we used Agreement Rate and Cohen's Kappa as the evaluation metrics.

\section{Conclusion and Future Work}
\label{sec:conclusion}

In this study, we first conducted a comprehensive empirical study of LLM-as-Judge methods, identifying key differences between general and reasoning models.
We then proposed {\tool}, a code evaluation metric balancing accuracy, efficiency, and explainability. 
Our experiments confirm that {\tool} outperforms state-of-the-art methods while maintaining robustness to preference leakage.

Future work will focus on: 
(1) evaluating {\tool} across additional programming languages and more complex datasets to better understand its generalization capabilities, and 
(2) applying {\tool} as an environment for reinforcement learning to improve code generation models.




\bibliographystyle{IEEEtran}
\bibliography{mylib}
\end{document}